\documentclass{sig-alternate-2013}
\toappear{To appear in the fifth International Conference on Future Energy Systems (ACM e-Energy), Cambridge, UK. 2014}

\usepackage[hidelinks]{hyperref}
\usepackage{comment}
\usepackage{array}
\usepackage{url}
\usepackage{listings}
\usepackage{algorithm}
\usepackage{algorithmic}

\newcommand{\figref}[1]{Figure~\ref{#1}}
\newcommand{\secref}[1]{Section~\ref{#1}}
\newcommand{\appref}[1]{Appendix~\ref{#1}}
\newcommand{\tabref}[1]{Table~\ref{#1}}
\newcommand{\algoref}[1]{Algorithm~\ref{#1}}

\usepackage[font=small,labelfont=bf,textfont=bf]{caption}
\newcommand{\specialcell}[2][c]{%
  \begin{tabular}[#1]{@{}c@{}}#2\end{tabular}}

\usepackage{xcolor}
\usepackage{colortbl}
\usepackage{xfrac}

\title{NILMTK: An Open Source Toolkit for Non-intrusive Load Monitoring}

\vspace{-2pt}
\author{Nipun Batra$^1$, Jack Kelly$^2$, Oliver Parson$^3$, Haimonti Dutta$^4$, William Knottenbelt$^2$,\\ Alex Rogers$^3$, Amarjeet Singh$^1$, Mani Srivastava$^5$\\ 
\small$^1$Indraprastha Institute of Information Technology Delhi, India ~\{nipunb,~amarjeet\}@iiitd.ac.in\\
\small$^2$ Imperial College London ~\{jack.kelly,~w.knottenbelt\}@imperial.ac.uk\\
\small$^3$ University of Southampton ~\{osp,~acr\}@ecs.soton.ac.uk\\
\small$^4$ CCLS Columbia ~\{haimonti@ccls.columbia.edu\}\\
\small$^5$ UCLA ~\{mbs@ucla.edu\}\\
}

\date{January 2014}

\begin{document}

\maketitle


\begin{abstract}
\noindent
\textit{
Non-intrusive load monitoring, or energy disaggregation, aims to separate household energy consumption data collected from a single point of measurement into appliance-level consumption data. In recent years, the field has rapidly expanded due to increased interest as national deployments of smart meters have begun in many countries. However, empirically comparing disaggregation algorithms is currently virtually impossible. This is due to the different data sets used, the lack of reference implementations of these algorithms and the variety of accuracy metrics employed. To address this challenge, we present the Non-intrusive Load Monitoring Toolkit~(NILMTK); an open source toolkit designed specifically to enable the comparison of energy disaggregation algorithms in a reproducible manner. This work is the first research to compare multiple disaggregation approaches across multiple publicly available data sets. Our toolkit includes parsers for a range of existing data sets, a collection of preprocessing algorithms, a set of statistics for describing data sets, two reference benchmark disaggregation algorithms and a suite of accuracy metrics. We demonstrate the range of reproducible analyses which are made possible by our toolkit, including the analysis of six publicly available data sets and the evaluation of both benchmark disaggregation algorithms across such data sets.
}
\end{abstract}




\section{Introduction}

\noindent
Non-intrusive load monitoring (NILM), or energy disaggregation, aims to break down a household's aggregate electricity consumption into individual appliances~\cite{hart_1992}. The motivations for such a process are threefold. First, informing a household's occupants of how much energy each appliance consumes empowers them to take steps towards reducing their energy consumption~\cite{darby_2006}. Second, personalised feedback can be provided which quantifies the savings of certain appliance-specific advice, such as the financial savings when an old inefficient appliance is replaced by a new efficient appliance. Third, if the NILM system is able to determine the time of use of each appliance, a recommender system would be able to inform the household's occupants of the savings of deferring appliance use to a time of day when electricity is either cheaper or has a lower carbon footprint.

Such benefits have drawn significant interest in the field since its inception 25 years ago. In recent years, the combination of smart meter meter deployments~\cite{CaliforniaPublicUtilitiesCommission2006,DepartmentofEnergy&ClimateChange2013} and reduced hardware costs of household electricity sensors has led to a rapid expansion of the field. Such rapid growth over the past five years has been evidenced by the wealth of academic papers published, international meetings held (e.g.\ NILM 2012\footnote{\url{http://www.ices.cmu.edu/psii/nilm/}} and EPRI NILM 2013\footnote{\url{http://goo.gl/dr4tpq}}), startup companies founded (e.g. Bidgely and Neurio) and data sets released, (e.g.\ REDD~\cite{redd}, BLUED~\cite{blued} and Smart*~\cite{smart}).

However, three core obstacles currently prevent the direct comparison of state-of-the-art approaches, and as a result may be impeding progress within the field. To the best of our knowledge, each contribution to date has only been evaluated on a single data set and consequently it is hard to assess whether such approaches generalise to new households. Furthermore, many researchers sub-sample data sets to select specific households, appliances and time periods, making experimental results more difficult to reproduce. Second, newly proposed approaches are rarely compared against the same benchmark algorithms, further increasing the difficulty in empirical comparisons of performance between different publications. Moreover, the lack of reference implementations of these state-of-the-art algorithms often leads to the reimplementation of such approaches. Third, many papers target different use cases for NILM and therefore the accuracy of their proposed approaches are evaluated using a different set of performance metrics. As a result the numerical performance calculated by such metrics cannot be compared between any two papers. These three obstacles have led to the proposal of successive extensions to state-of-the-art algorithms, while a direct comparison between new and existing approaches remains impossible.

Similar obstacles have arisen in other research fields and prompted the development of toolkits specifically designed to support research in that area. For example, PhysioToolkit offers access to over 50 databases of physiological data and provides software to support the processing and analysis of such data for the biomedical research community~\cite{physionet}. Similarly, CRAWDAD collects 89 data sets of wireless network data in addition to software to aid the analysis of such data for the wireless network community~\cite{crawdad}. However, no such toolkit is available to the NILM community.

Against this background, we propose NILMTK\footnote{Code: \url{http://github.com/nilmtk/nilmtk} (release v0.1.0 was used for this paper)}; an open source toolkit designed specifically to enable easy access to and comparative analysis of energy disaggregation algorithms across diverse data sets. NILMTK provides a complete pipeline from data sets to accuracy metrics, thereby lowering the entry barrier for researchers to implement a new algorithm and compare its performance against the current state of the art. NILMTK has been:
\begin{itemize}
\item released as open source software (with documentation\footnote{Documentation: \url{http://nilmtk.github.io/nilmtk}}) in an effort to encourage researchers to contribute data sets, benchmark algorithms and accuracy metrics as they are proposed, with the goal of enabling a greater level of collaboration within the community. 
\item designed using a modular structure, therefore allowing researchers to reuse or replace individual components as required. The API design is influenced by \texttt{scikit-learn}~\cite{scikit}, which is a machine learning library in Python, well known for its consistent API and complete documentation.
\item written in Python with flat file input and output formats, in addition to high performance binary formats, ensuring compatibility with existing algorithms written in any language and designed for any platform.
\end{itemize}

\begin{table*}[]
  \centering
  \begin{tabular}{c c c c c c c c}
    \hline
    \bf  & \bf  & \bf  & \bf Duration & \bf Number & \bf Appliance & \bf Aggregate\\
    \bf Data set & \bf Institution & \bf Location & \bf per  & \bf of & \bf sample & \bf sample\\
    \bf  & \bf  & \bf  & \bf house & \bf houses & \bf frequency & \bf frequency \\
    \hline
    REDD (2011) & MIT & MA, USA & 3-19 days & 6 & 3 sec & 1 sec \& 15 kHz\\
    BLUED (2012) & CMU & PA, USA & 8 days & 1 & N/A* & 12 kHz\\
    Smart* (2012) & UMass & MA, USA & 3 months & 3 & 1 sec & 1 sec\\
    Tracebase (2012) & Darmstadt & Germany & N/A & N/A & 1-10 sec & N/A\\
    Sample (2013) & Pecan Street & TX, USA & 7 days & 10 & 1 min & 1 min\\
    HES (2013) & DECC, DEFRA & UK & 1 or 12 months & 251 & 2 or 10 min
    & 2 or 10 min\\
    AMPds (2013) & Simon Fraser U. & BC, Canada & 1 year & 1 & 1 min & 1 min\\
    iAWE (2013) & IIIT Delhi & Delhi, India & 73 days & 1 & 1 or 6 sec & 1 sec\\
    UK-DALE (2014) & Imperial College & London, UK & 3-17 months & 4 & 6 sec & 1-6 sec \& 16 kHz\\
    \hline
  \end{tabular}
  \caption{Comparison of household energy data sets. *BLUED labels state transitions for each appliance.}
  \label{table:datasets}
\end{table*}

The contributions of NILMTK are summarised as follows:
\begin{itemize}
\item We propose NILMTK-DF (data format), the standard energy disaggregation data structure used by our toolkit.  NILMTK-DF is modelled loosely on the REDD data set format~\cite{redd} to allow easy adoption within the community. Furthermore, we provide parsers from six existing data sets into our proposed NILMTK-DF format. 
\item We provide statistical and diagnostic functions which provide a detailed understanding of each data set.  We also provide preprocessing functions for mitigating common challenges with NILM data sets.
\item We provide implementations of two benchmark disaggregation algorithms: first an approach based on combinatorial optimisation~\cite{hart_1992}, and second an approach based on the factorial hidden Markov model~\cite{redd,kim_2011}. We demonstrate the ease by which NILMTK allows the comparison of these algorithms across a range of existing data sets, and present results of their performance.
\item We present a suite of accuracy metrics which enables the evaluation of any disaggregation algorithm compatible with NILMTK. This allows the performance of a disaggregation algorithm to be evaluated for a range of use cases.
\end{itemize}

The remainder of this paper is organised as follows. In \secref{sec:related} we provide an overview of related work. In \secref{sec:nilmtk} we present NILMTK and describe its components. In \secref{evaluation} we demonstrate the empirical evaluations which are enabled by NILMTK, and provide analyses of existing data sets and disaggregation algorithms. Finally, in \secref{sec:conclusions} we conclude the paper and propose directions for future work.

\section{Background}
\label{sec:related}

\noindent
The field of non-intrusive load monitoring was founded 25 years ago when Hart proposed the first algorithm for the disaggregation of household energy usage~\cite{hart_1992,armel_2013}. 
However, the majority of research prior to 2011 had been evaluated using either lab-based or simulated data and hence the performance of disaggregation algorithms in real households had remained unknown. More recently, national deployments of smart meters have prompted a renewed interest in energy disaggregation. 
We now discuss recent research which has contributed new data sets (\secref{sec:datasets}), disaggregation algorithms (\secref{sec:algorithms}) and evaluation metrics (\secref{sec:evaluation_metrics}) to the field. 
In \secref{sec:need_nilm_toolkit} we discuss general purpose toolkits, and finally in \secref{sec:notation} we formalise the NILM problem drawing upon notation used in prior literature.

\subsection{Public Data Sets}
\label{sec:datasets}
\noindent In 2011, the Reference Energy Disaggregation Dataset (REDD) \cite{redd} was introduced as the first publicly available data set collected specifically to aid NILM research. The data set contains both aggregate and sub-metered power data from six households, and has since become the most popular data set for evaluating energy disaggregation algorithms. In 2012, the Building-Level fUlly-labeled dataset for Electricity Disaggregation (BLUED) \cite{blued} was released containing data from a single household. However, the data set does not include sub-metered power data, and instead records events triggered by appliance state changes. 
As a result, it is only possible to evaluate whether changes in appliance states have been detected (e.g.\ washing machine turns on), rather than the assignment of aggregate power demand to individual appliances (e.g.\ washing machine draws 2~kW power). More recently, the Smart*~\cite{smart} data set was released, which contains household aggregate power data from three households, while sub-metered appliance power data was only collected from a single household.

In 2013 the Pecan Street sample data set was released~\cite{pecan}, which contains both aggregate and sub-metered power data from 10 households. Later, the Household Electricity Survey data set was released~\cite{hes}, which contains data from 251 households although aggregate data was only collected for 14 households. The Almanac of Minutely Power dataset (AMPds)~\cite{ampds} was also released that year containing both aggregate and sub-metered power data from a single household. Subsequently, the Indian data for Ambient Water and Electricity Sensing (iAWE)~\cite{iawe} was released, which contains both aggregate and sub-metered power data from a single house. Most recently, the UK Domestic Appliance-Level Electricity data set~\cite{UK-DALE} (UK-DALE) was released which contains data from four households using both aggregate meters and individual appliance sub-meters. Unfortunately, subtle differences in the aims of each data set have led to completely different data formats being used. As a result, a time-consuming engineering barrier exists when using the data sets, each of which are in different formats. This has resulted in publications using only a single data set to evaluate a given approach, and consequently the generality of results over large numbers of households are rarely investigated. We summarise these data sets in \tabref{table:datasets}.

\begin{figure*}
\centering \includegraphics[scale=0.7]{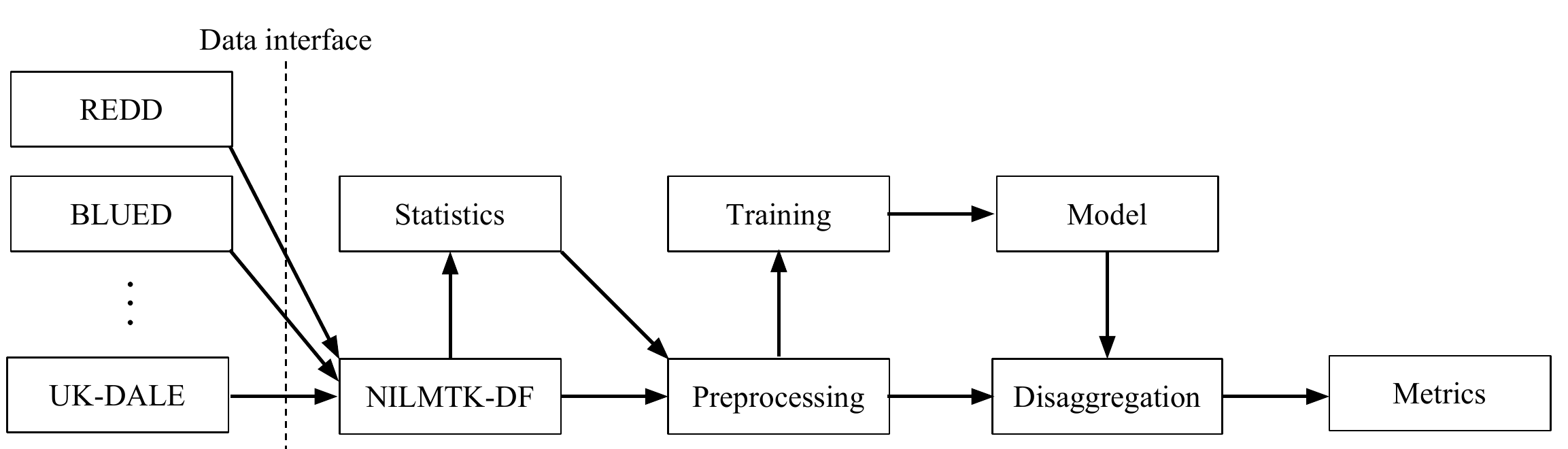}
\caption{NILMTK pipeline. At each stage of the pipeline, results and data can be stored to or loaded from disk.}
\label{fig:pipeline}
\end{figure*}

\medskip 

\subsection{Disaggregation Algorithms \& Benchmarks}
\label{sec:algorithms}
\noindent The REDD data set was proposed along with a performance result of a benchmark disaggregation algorithm using 10 second data across five of the six households~\cite{redd}. Kolter and Jaakkola later proposed an extension to the benchmark algorithm~\cite{kolter_2012}, however the extension was only evaluated using features extracted from 15~kHz data from a single house from the data set, and therefore the performance results are not directly comparable. Later, Zeifman~\cite{zeifman_2012} and Johnson and Willsky~\cite{johnson_2013} evaluated various approaches using the same data set, although both selected a different subset of appliances and calculated an artificial household aggregate from these appliances, therefore simplifying the disaggregation problem and preventing a numerical comparison with other publications. Subsequently, Parson et al.~\cite{parson_2012} and Rahayu et al.~\cite{rahayu_2012} both proposed new approaches, although each were evaluated using a different set of four houses from the REDD data set, again preventing a numerical comparison between publications. Last, Batra et al.~\cite{batra_2013} evaluated their approach on the REDD data set using a different household to Kolter and Jaakkola. As a result, it has not been possible to deduce whether one approach is preferable to another from the literature.

The BLUED data set was introduced along with a benchmark algorithm~\cite{blued}, but has since only been used by one other publication~\cite{anderson_2012}. Similarly, AMPds has only been used to evaluate disaggregation algorithms proposed by the data set authors~\cite{ampds}. Clearly, the variety of different formats is slowing the uptake of new data sets, and also preventing algorithms from being tested across multiple data sets. 

It is essential to compare newly proposed disaggregation algorithms to the state of the art in order to assess the increase in an algorithm's performance. However, the lack of available reference implementations of state-of-the-art disaggregation algorithms has led to authors often comparing against more basic benchmark algorithms. This problem is further compounded since there is no single consensus on which benchmarks to use, and as a result most publications use a different benchmark algorithm. For example, Kolter and Jaakkola compared their approach to a set of decoupled HMMs~\cite{kolter_2012}, Parson et al.\ and Batra et al.\ both evaluated their approaches against variants of their own approaches~\cite{parson_2012,batra_2013}, Zeifman compared their approach to a Bayesian classifier, while Rahayu et al. and Johnson and Willsky both compared against a factorial hidden Markov model (FHMM)~\cite{rahayu_2012,johnson_2013}. Clearly, further publications would benefit from openly available benchmark algorithms against which newly proposed algorithms could be easily compared.

\medskip 

\subsection{Evaluation Metrics}
\label{sec:evaluation_metrics}
\noindent The range of different application areas of energy disaggregation has prompted a number of evaluation metrics to be proposed. For example, four disaggregation metrics labelled \emph{energy correctly assigned} have recently been used to evaluate the performance of disaggregation algorithms using the REDD data set. First, Kolter and Johnson~\cite{redd} proposed an accuracy metric which captures the error in assigned energy normalised by the actual energy consumption in each time slice averaged over all appliances, which was also later used by Rahayu et al.~\cite{rahayu_2012} and Johnson and Willsky~\cite{johnson_2013}. However, large errors in the assigned energy in some time slices will result in a negative accuracy, making this an ill-posed metric. Second, Kolter and Jaakkola~\cite{kolter_2012} proposed an equivalent metric wherein the error is presented individually for each appliance rather than an average across all appliances. Third, Parson et al.~\cite{parson_2012} proposed a metric which captures the error in assigned energy consumed over the complete duration of the data set rather than per time slice. This metric allows overestimates and underestimates in the assigned energy in different time slices to cancel out, and therefore does not represent all disaggregation errors. Fourth, Batra et al.~\cite{batra_2013} proposed a subtly different metric to Kolter and Johnson~\cite{redd}, in which error is reported instead of accuracy, and also energy assigned to an incorrect appliance is double counted as both an overestimate of one appliance's energy consumption and an underestimate of another. The differences between these four metrics prevent numerical comparisons between publications, and motivate the use of common metrics.

\subsection{General Purpose Toolkits}
\label{sec:need_nilm_toolkit}
\noindent Although no toolkit currently exists specifically for energy disaggregation, various toolkits are available for more general machine learning tasks. For example, \texttt{scikit-learn} is a general purpose machine learning toolkit implemented in Python~\cite{scikit} and \texttt{GraphLab} is a machine learning and data mining toolkit written in C++~\cite{graphlab}. While such toolkits provide generic implementations of machine learning algorithms, they lack functionality specific to the energy disaggregation domain, such as data set parsers, benchmark disaggregation algorithms, and energy disaggregation metrics. Therefore, an energy disaggregation toolkit should extend such general toolkits rather than replace them, in a similar way that \texttt{scikit-learn} adds machine learning functionality to the \texttt{numpy} numerical library for Python. 

\medskip 

\subsection{Energy Disaggregation Definition}
\label{sec:notation}
\noindent The aim of energy disaggregation is to provide estimates, $\hat{y}^{(n)}_t$, of the actual power demand, $y^{(n)}_t$, of each appliance $n$ at time $t$, from household aggregate power readings, $\bar{y}_t$. Most NILM algorithms model appliances using a set of discrete states such as off, on, intermediate, etc.  We use $x^{(n)}_t \in \mathbb Z_{> 0}$ to represent the ground truth state, and $\hat{x}^{(n)}_t$ to represent the appliance state estimated by a disaggregation algorithm.

\medskip 

\section{NILMTK}
\label{sec:nilmtk}

\noindent
We designed NILMTK with two core use cases in mind. First, it should enable the analysis of existing data sets and algorithms. Second, it should provide a simple interface for the addition of new data sets and algorithms.
To do so, we implemented NILMTK in Python due to the availability of a vast set of libraries supporting both machine learning research (e.g.\ \texttt{Pandas}, \texttt{scikit-learn}) and the deployment of such research as web applications (e.g.\ \texttt{Django}). Furthermore, Python allows easy deployment in diverse environments including academic settings and is increasingly being used for data science.


\figref{fig:pipeline} presents the NILMTK pipeline from the import of data sets to the evaluation of various disaggregation algorithms over various metrics. In \appref{app:appendix_pipeline} we summarise the NILMTK pipeline with an illustrative code snippet. In the remainder of this section we discuss each module of the pipeline: the NILMTK data format, the data set diagnostics and statistics, preprocessing, disaggregation, model import and export and finally we describe accuracy metrics.

\medskip 

\subsection{Data Format}
\label{sec:data_format} 

\noindent
Motivated by our discussion in \secref{sec:datasets} of the wide differences between multiple data sets released in the public domain, we propose NILMTK-DF; a common
data set format inspired by the REDD format~\cite{redd}, into which
existing data sets can be converted. NILMTK currently includes
importers for the following six data sets: REDD, Smart*, Pecan Street, iAWE, AMPds
and UK-DALE. BLUED was excluded due to the lack of sub-metered power data, the Tracebase data set was excluded due to the lack of household aggregate power data and HES was excluded due to time constraints.

After import, the data resides in our NILMTK-DF in-memory data structure, which is used throughout the NILMTK pipeline. Data can be saved or loaded from disk at multiple stages in the NILMTK processing pipeline to allow other tools to interact with NILMTK.  We provide two CSV flat file formats: a rich NILMTK-DF CSV format and a ``strict REDD" format which allows researchers to use their existing tools designed to process REDD data.  We also provide a more efficient binary format using the Hierarchical Data Format (HDF5).  In addition to storing electricity data, NILMTK-DF can also store relevant metadata and other sensor modalities such as gas, water, temperature, etc. It has been shown that such additional sensor and metadata information may help enhance NILM prediction~\cite{schoofs_2010}. 

Another important feature of our format is the standardisation of nomenclature.  Different data sets use different labels for the same class of appliance (e.g.\ REDD uses `refrigerator' whilst AMPds uses `FGE') and different names for the measured parameters.  When data is first imported into NILMTK, these diverse labels are converted to a standard vocabulary~\cite{NILM_Metadata}.

In addition, NILMTK allows rich metadata to be associated with a
household, appliance or meter.  For example, NILMTK can store the
parameters measured by each meter (e.g.\ reactive power, real power),
the geographical coordinates of each house (to enable weather data to be retrieved), the mains wiring
defining the meter hierarchy (useful if a single appliance is measured
at the appliance, circuit and aggregate levels), whether a single
meter measures multiple appliances and whether a specific lamp is
dimmable. More detail is provided in
\appref{app:appendix_data_format} and our full NILM Metadata schema is described in~\cite{NILM_Metadata}.

Through such a combination of metadata and standard nomenclature, NILMTK allows for analysis of appliance data across multiple data sets. For example, users can perform queries such as:
`what is the energy consumption of refrigerators in the USA
compared to the UK?'. Further examples are given in~\appref{app:examples}.

We have defined a common interface for data set importers
which, combined with the definition of our in-memory data structures,
enables developers to easily add new data set
importers to NILMTK.

\subsection{Data Set Diagnostics}
\label{sec:diagnostic_definitions}

\noindent
Since no data set is perfect, researchers are required to explore the characteristics of each data set before
disaggregation approaches can be evaluated.  To help diagnose these issues, NILMTK
provides diagnostic functions including:

\textbf{Detect gaps:} Many NILM algorithms assume that each sensor channel is
contiguous. However, this assumption is violated when sensors are off or
malfunctioning.  A `gap' exists between any pair of consecutive
samples if the time elapsed between them is larger than
a predefined threshold.

\textbf{Dropout rate:} The
dropout rate is the total number of recorded samples, divided by the
number of expected samples (which is the length of the time window
under consideration multiplied by the sample rate).

\textbf{Dropout rate (ignoring gaps):} To quantify the rate
at which a wireless sensor drops samples due to radio issues, we
 first remove large gaps where the sensor is off and subsequently
 calculate the dropout rate for the remaining contiguous sections.

\textbf{Up-time:}  The up-time is the total time for which a sensor
was recording.  It is the last timestamp, minus the first timestamp,
minus the duration of any gaps.

\textbf{Diagnose:} NILMTK provides a single \texttt{diagnose}
function which checks for all the issues we have encountered.

\subsection{Data Set Statistics}
\label{sec:stats_definitions}

\noindent
Distinct from \emph{diagnostic} statistics, NILMTK also provides
functions for exploring appliance usage, e.g.:

\textbf{Proportion of energy sub-metered:} Data sets rarely sub-meter
every appliance or circuit, and as a result it is useful to quantify the proportion of
total energy measured by sub-metered channels. 
Prior to calculating this
statistic, all gaps present in the mains recordings are masked out of
each sub-metered channel, and therefore any additional missing sub-meter data is assumed to
be due to the meter and load being switched off.

\secref{sec:diagnostic_definitions} and \ref{sec:stats_definitions} have described a subset of the diagnostic and
statistical functions in NILMTK.  Further functions are listed
in Appendix~\ref{app:functions} and in the statistics section of the
online documentation.\footnote{
  \url{http://nilmtk.github.io/nilmtk/stats.html}}

\subsection{Preprocessing of Data Sets}
\label{sec:preprocessing}

\noindent
To mitigate the problems  with different data sets, some of which were presented in \secref{sec:diagnostic_definitions}, NILMTK provides several preprocessing functions, including:

\textbf{Downsample:} As seen in \tabref{table:datasets}, the sampling rate of appliance monitors varies
from 0.008~Hz to 16~kHz across the data sets. The downsample preprocessor down-samples data sets to a specified frequency using aggregation functions such as mean, mode and median.

\textbf{Voltage normalisation:} The data sets presented in \tabref{table:datasets} have been
collected from different countries, where voltage fluctuations vary
widely. Batra et al. showed voltage fluctuates from 180-250~V in the
iAWE data set collected in India~\cite{iawe}, while the voltage in the
Smart* data set varies across the range 118-123~V. Hart suggested to account for these voltage fluctuations as they can
significantly impact power draw~\cite{hart_1992}. Therefore, NILMTK
provides a voltage normalisation function based on Hart's equation:
\begin{equation}
\textit{Power}_{\textit{normalised}} = 
\left(\frac{\textit{Voltage}_{\textit{nominal}}}{\textit{Voltage}_{\textit{observed}}}\right)^2 
\times \textit{Power}_{\textit{observed}}
\end{equation}

\textbf{Top-$k$ appliances}: It is often advantageous to model the top-$k$ energy consuming appliances instead of all appliances for the following three reasons. 
First, the disaggregation of such appliances provides the most value. Second, such appliances contribute the most salient features, and therefore the remaining appliances can be considered to contribute only noise. Third, each additional modelled appliance might contribute significantly to the complexity of the disaggregation task. Therefore, NILMTK provides a function to identify the top-$k$ energy consuming appliances.


NILMTK also provides preprocessing functions for fixing
other common issues with these data sets, such as: (i) interpolating
small periods of missing data when appliance sensors did not report
readings, (ii) filtering out implausible values (such as readings
where observed voltage is more than twice the rated voltage) and (iii)
filtering out appliance data when mains data is missing.

Each data set importer defines a \texttt{preprocess}
function which runs the necessary preprocessing functions to clean
the specific data set. A detailed account of preprocessing functions supported by NILMTK can be found in
\appref{app:functions} and in the online documentation.\footnote{
  \url{http://nilmtk.github.io/nilmtk/preprocessing.html}}

\subsection{Training and Disaggregation Algorithms}
\label{sec:training}
\noindent
NILMTK provides implementations of two common benchmark disaggregation algorithms: combinatorial optimisation (CO) and factorial hidden Markov model (FHMM). CO was proposed by Hart in his seminal work~\cite{hart_1992}, while techniques based on extensions of the FHMM have been proposed more recently~\cite{redd,kim_2011}. The aim of the inclusion of these algorithms is not to present state-of-the-art disaggregation results, but instead to enable new approaches to be compared to well-studied benchmark algorithms without requiring the reimplementation of such algorithms. We now describe these two algorithms.

\textbf{Combinatorial Optimisation:}
CO finds the optimal combination of appliance states, which minimises the difference between the sum of the predicted appliance power and the observed aggregate power, subject to a set of appliance models. 
\begin{equation}
\hat{x}^{(n)}_t=\operatorname*{arg min}_{\hat{x}^{(n)}_t}\left|\bar{y}_t-\sum\limits_{n=1}^{N}\hat{y}^{(n)}_t\right|
\end{equation}
Since each time slice is considered as a separate optimisation problem, each time slice is assumed to be independent.
CO resembles the subset sum problem and thus is NP-complete. The complexity of disaggregation for $T$ time slices is $O(TK^N)$, where $N$ is the number of appliances and $K$ is the number of appliance states. Since the complexity of CO is exponential in the number of appliances, the approach is only computationally tractable for a small number of modelled appliances.

\textbf{Factorial Hidden Markov Model:} The power demand of each appliance can be modelled as the observed value of a hidden Markov model (HMM). The hidden component of these HMMs are the states of the appliances. Energy disaggregation involves jointly decoding the power draw of $n$ appliances and hence a factorial HMM~\cite{fhmm} is well suited. A FHMM can be represented by an equivalent HMM in which each state corresponds to a different combination of states of each appliance. Such a FHMM model has three parameters: (i) prior probability ($\pi$) containing $K^N$ entries, (ii) transition matrix ($A$) containing $K^N \times K^N$ or $K^{2N}$ entries, and (iii) emission matrix ($B$) containing $2K^N$ entries. The complexity of exact disaggregation for such a model is $O(TK^{2N})$, and as a result FHMMs scale even worse than CO. From an implementation perspective, even storing (or computing) $A$ for 14 appliances with two states each consumes 8 GB of RAM. Hence, we propose to validate FHMMs on preprocessed data where the top-$k$ appliances are modelled, and appliances contributing less than a given threshold are discarded. However, it should be noted that more efficient pseudo-time algorithms could alternatively be used for inference over both CO and FHMM.

For algorithms such as FHMMs, it is necessary to model the relationships amongst consecutive samples. Thus, NILMTK provides facilities for dividing data into continuous sets for training and testing. While we have discussed supervised and non-event based algorithms here, NILMTK also supports event based and unsupervised approaches. Details for adding new algorithms are provided in \appref{app:new_algo}.

\subsection{Appliance Model Import and Export}


\noindent
Many approaches require sub-metered power data to be collected for training purposes from the same household in which disaggregation is to be performed. However, such data is costly and intrusive to collect, and therefore is unlikely to be available in a large-scale deployment of a NILM system. As a result, recent research has proposed training methods which do not require sub-metered power data to be collected from each household~\cite{kim_2011,parson_2012}. To provide a clear interface between training and disaggregation algorithms, NILMTK provides a \emph{model} module which encapsulates the results of the training module required by the disaggregation module. Each implementation of the module must provide import and export functions to interface with a JSON file for persistent model storage. NILMTK currently includes importers and exporters for both the FHMM and CO approaches described in \secref{sec:training}.


\begin{table*}[]
  \centering
\begin{tabular}{cccccc}
\hline
\textbf{Data set} & \textbf{\specialcell[h]{Number of\\appliances}} & \textbf{\specialcell[h]{Percentage\\energy\\sub-metered}} & \textbf{\specialcell[h]{Dropout rate\\(percent)\\ignoring gaps}} & \textbf{\specialcell[h]{Mains up-time\\per house\\(days)}} & \textbf{\specialcell[h]{Percentage\\up-time}} \\
\hline
REDD & 9, 16, 23 &58, 71, 89 & 0, 10, 16 &4, 18, 19 & 8, 40, 79 \\
Smart* &25 & 86 & 0 & 88 & 96 \\
Pecan Street &13, 14, 22 & 75, 87, 150 & 0, 0, 0 &7, 7, 7 & 100, 100, 100 \\
AMPds&20 &97 & 0 &364 & 100 \\
iAWE &10 &48 & 8 & 47 &93 \\
UK-DALE & 4, 12, 53 &19, 48, 82 &0, 7, 22 & 36, 102, 470 & 73, 84, 100 \\
\hline
\end{tabular}
  \caption{Summary of data set results calculated by the diagnostic and statistical
    functions in NILMTK.  Each cell represents the range of values
    across all households per data set.  The three
    numbers per cell are the minimum, median and maximum values. AMPds, Smart*
    and iAWE each contain just a single house, hence these rows
    have a single number per cell.}
  \label{table:dataset_results}
\end{table*}

\subsection{Accuracy Metrics}
\label{sec:metrics}

\noindent
As discussed in \secref{sec:evaluation_metrics}, a range of accuracy metrics are required due to the diversity of application areas of energy disaggregation research. To satisfy this requirement, NILMTK provides a set of metrics which combines both general detection metrics and those specific to energy disaggregation. We now give a brief description of each metric implemented in NILMTK along with its mathematical definition.

\textbf{Error in total energy assigned:} The difference between the total assigned energy and the actual energy consumed by appliance $n$ over the entire data set.
\begin{equation}
        \left | \sum_t y^{(n)}_t - \sum_t \hat{y}^{(n)}_t \right |
\end{equation}

\textbf{Fraction of total energy assigned correctly:} The overlap between the fraction of energy assigned to each appliance and the actual fraction of energy consumed by each appliance over the data set.
\begin{equation}
        \sum_n \mathrm{min} \left ( 
        \frac{\sum_n y^{(n)}_t}{\sum_{n,t} y^{(n)}_t}, 
        \frac{\sum_n \hat{y}^{(n)}_t}{\sum_{n,t} \hat{y}^{(n)}_t} 
        \right )
\end{equation}

\textbf{Normalised error in assigned power:} The sum of the differences between the assigned power and actual power of appliance $n$ in each time slice $t$, normalised by the appliance's total energy consumption.
\begin{equation}
        \frac
        { \sum_t {\left | y_t^{(n)} - \hat{y}_t^{(n)} \right |} }
        { \sum_t y_t^{(n)} }
\end{equation}

\textbf{RMS error in assigned power:} The root mean square error between the assigned power and actual power of appliance $n$ in each time slice $t$.
\begin{equation}
\sqrt{ \frac{1}{T} \sum_t{ \left ( y^{(n)}_t - \hat{y}^{(n)}_t \right )^2 } }
\end{equation}

\textbf{Confusion matrix:} The number of time slices in which each of an appliance's states were either confused with every other state or correctly classified.

\textbf{True positives, False positives, False negatives, True negatives:} The number of time slices in which appliance $n$ was either correctly classified as being on~($\mathit{TP}$), classified as being on while it was actually off~($\mathit{FP}$), classified as off while is was actually on~($\mathit{FN}$) and correctly classified as being off~($\mathit{TN}$).
\begin{equation}
\mathit{TP}^{(n)} = 
\sum_{t}
\mathrm{AND} \left ( x^{(n)}_t = \mathit{on}, \hat{x}^{(n)}_t = \mathit{on} \right )
\end{equation}
\begin{equation}
\mathit{FP}^{(n)} = 
\sum_{t}
\mathrm{AND} \left ( x^{(n)}_t = \mathit{off}, \hat{x}^{(n)}_t = \mathit{on} \right )
\end{equation}
\begin{equation}
\mathit{FN}^{(n)} = 
\sum_{t}
\mathrm{AND} \left ( x^{(n)}_t = \mathit{on}, \hat{x}^{(n)}_t = \mathit{off} \right )
\end{equation}
\begin{equation}
\mathit{TN}^{(n)} = 
\sum_{t}
\mathrm{AND} \left ( x^{(n)}_t = \mathit{off}, \hat{x}^{(n)}_t = \mathit{off} \right )
\end{equation}

\textbf{True/False positive rate:} The fraction of time slices in which an appliance was correctly predicted to be on that it was actually on~($\mathit{TPR}$), and the fraction of time slices in which the appliance was incorrectly predicted to be on that it was actually off~($\mathit{FPR}$). We omit appliance indices $n$ in the following metrics for clarity.
\begin{equation}
\mathit{TPR} = \frac{\mathit{TP}}{\left ( \mathit{TP} + \mathit{FN} \right )}
\end{equation}
\begin{equation}
\mathit{FPR} = \frac{\mathit{FP}}{\left ( \mathit{FP} + \mathit{TN} \right )}
\end{equation}

\textbf{Precision, Recall:} The fraction of time slices in which an appliance was correctly predicted to be on that it was actually off~(Precision), and the fraction of time slices in which the appliance was correctly predicted to be on that it was actually on~(Recall).
\begin{equation}
\mathit{Precision} = \frac{\mathit{TP}}{\left ( \mathit{TP} + \mathit{FP} \right )}
\end{equation}
\begin{equation}
\mathit{Recall} = \frac{\mathit{TP}}{\left ( \mathit{TP} + \mathit{FN} \right )}
\end{equation}

\textbf{F-score:} The harmonic mean of precision and recall.
\begin{equation}
\mathit{F\text{-}score} = \frac
            {2 . \mathit{Precision} . \mathit{Recall}}
            {\mathit{Precision} + \mathit{Recall}}
\end{equation}

\textbf{Hamming loss:} The total information lost when appliances are incorrectly classified over the data set.
\begin{equation}
\mathit{HammingLoss} = 
        \frac{1}{T} \sum_{t}
        \frac{1}{N} \sum_{n}
        \mathrm{XOR} \left ( x^{(n)}_t, \hat{x}^{(n)}_t \right )
\end{equation}

\section{Evaluation}
\label{evaluation}

\noindent
We now demonstrate several examples of the rich analyses supported by NILMTK. First, we diagnose some common (and inevitable) issues in a selection of data sets.  Second, we show various patterns of appliance usage. Third, we give some examples of the effect of voltage normalisation on the power demand of individual appliances, and discuss how this might affect the performance of a disaggregation algorithm. Fourth, we present summary performance results of the two benchmark algorithms included in NILMTK across six data sets using a number of accuracy metrics. Finally, we present detailed results of these algorithms for a single data set, and discuss their performance for different appliances.

\subsection{Data Set Diagnostics}

\begin{figure}[!t]
  \centering
  \includegraphics{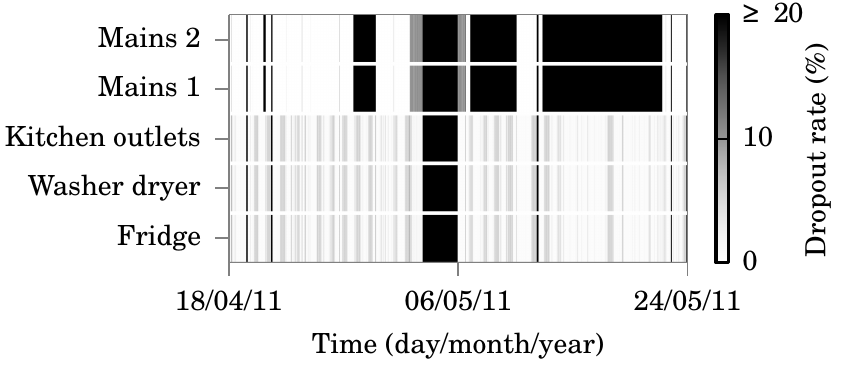} 
  \caption{Lost samples per hour from a representative subset of channels in REDD house 1.}
  \label{fig:lost_samples} 
\end{figure}

\noindent
\tabref{table:dataset_results} shows a selection of diagnostic and statistical functions 
(defined in Section~\ref{sec:diagnostic_definitions} and~\ref{sec:stats_definitions}) computed by NILMTK
across six public data sets. BLUED, Tracebase and HES were not included for the same reasons as in \secref{sec:data_format}. The table illustrates
that AMPds used a robust recording platform because it has a
percentage up-time of 100\%, a dropout rate of zero and 97\% of the
energy recorded by the mains channel was captured by the sub-meters.
Similarly, Pecan Street has an up-time of 100\% and zero dropout rate.  However, two homes
in the Pecan Street data registered a proportion of energy sub-metered of over
100\%. This indicates that some overlap exists between the metered channels, and as a result some appliances are metered by multiple channels.
This illustrates the importance of data set metadata (proposed as part
of NILMTK-DF in Section~\ref{sec:data_format}) describing the basic
mains wiring.

Figure~\ref{fig:lost_samples} shows the distribution of missing
samples for REDD house 1.  From this we can see that each mains
recording channel has four large gaps (the solid black blocks) where
the sensors are off. The sub-metered channels have
only one large gap.  Ignoring this gap and focusing on the time
periods where the sensors are recording, we see numerous periods where
the dropout rate is around 10\%.  Such issues are by no means unique
to REDD and are crucial to diagnose before data sets can be used for
the evaluation of disaggregation algorithms or for data set
statistics.

\begin{figure}
  \centering
  \includegraphics[scale=1]{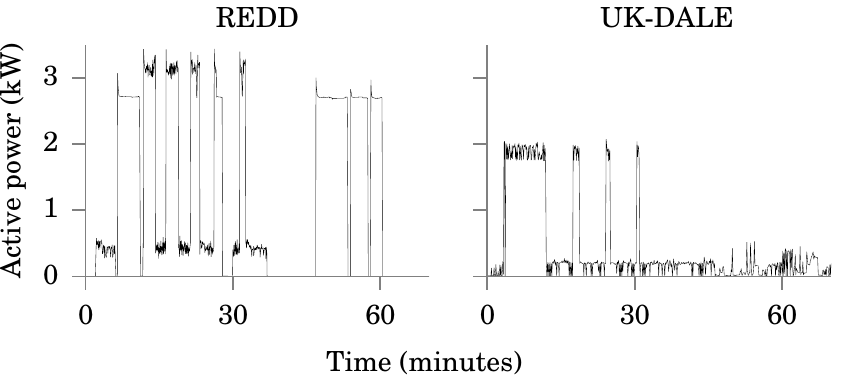}
  \caption{Comparison of power draw of washing machines in one house from REDD (USA) and
   UK-DALE.}
  \label{fig:wm}
\end{figure} 

\begin{figure*}
  \centering
  \includegraphics[scale=1]{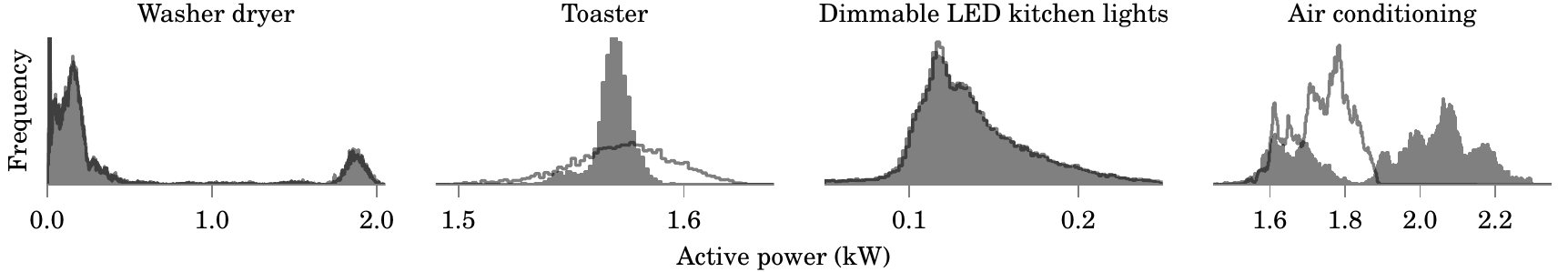} 
  \caption{Histograms of power consumption. The filled grey plots show
    histograms of normalised power.  The thin, grey,
    semi-transparent lines drawn over the filled plots show histograms
    of un-normalised power.}
  \label{fig:power_histograms} 
\end{figure*}

\subsection{Data Set Statistics}

\noindent
Energy disaggregation systems must model individual appliances.  Hence, as well as diagnosing technical issues with each data set, NILMTK also provides functions to visualise patterns of behaviour recorded in each data set. For example, different appliances draw a different amount of power (e.g.\ a toaster draws approximately 1.57~kW), are used at different times of day (e.g.\ the TV is usually on in the evening) and have different correlations with external factors such as weather (e.g.\ lower outside temperature implies more usage of electric heating). Furthermore, load profiles of different appliances of the same type can vary considerably, especially appliances from different countries (e.g.\ the two washing machine profiles in Figure~\ref{fig:wm}). Some disaggregation systems benefit by capturing these patterns (for example, the conditional factorial hidden Markov model (CFHMM)~\cite{kim_2011} can model the influence of time of day on appliance usage). In the following sections, we present examples of how such information can be extracted from existing data sets using NILMTK, covering the distribution of appliance power demands (\secref{sec:power_hist}), usage patterns (\secref{sec:usage_hist}) and external dependencies (\secref{sec:weather_correlation}).

\subsubsection{Appliance power demands}
\label{sec:power_hist}

\noindent
Figure~\ref{fig:power_histograms} displays histograms of the
distribution of powers used by a selection of appliances (the washer dryer, toaster and dimmable LED kitchen lights are from UK-DALE house 1; the air conditioning unit is from iAWE).  Appliances
such as toasters and kettles tend to have just two possible power states:
on and off.  This simplicity makes them amenable to be modelled by,
for example, Markov chains with only two states per chain.  In contrast, more complex appliances
such as washing machines, vacuum cleaners and computers often
have many more states.

\begin{figure}[t]
 \centering
 \includegraphics[width=\columnwidth]{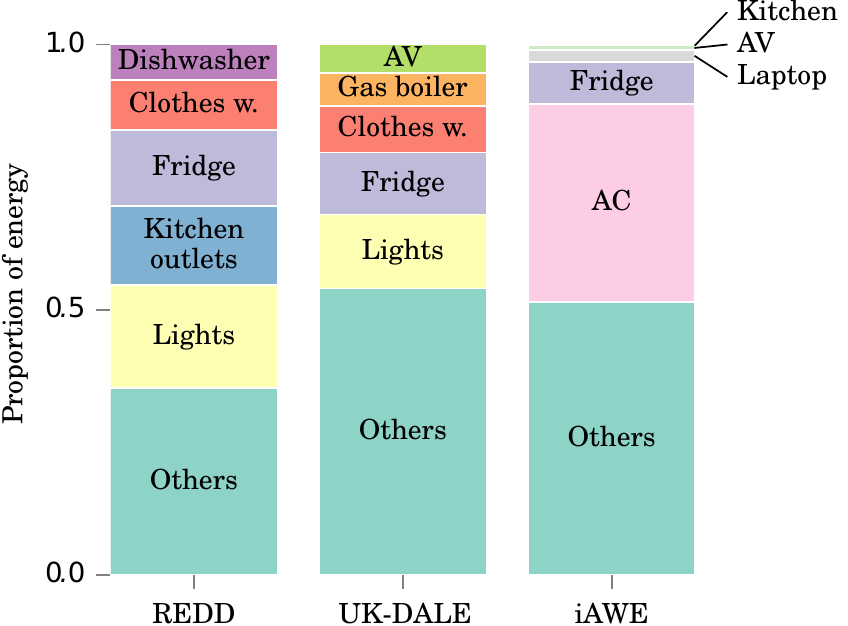}
 \caption{Top five appliances in terms of the proportion of the total
   energy used in a single house (house 1) in each of REDD (USA), iAWE (India) and
   UK-DALE.}
 \label{fig:pie}
\end{figure}

Figure~\ref{fig:pie} shows examples of how the proportion of energy use per appliance varies between countries. It can seen that the REDD and UK-DALE households share some similarities in the breakdown of household energy consumption. In contrast, the iAWE house shows a vastly different energy breakdown. For example, the house recorded in India for the iAWE
data set has two air conditioning units which account for almost half of the household's energy consumption, whilst the example household from the UK-DALE data set does not even contain an air conditioner.

%
%
%

\subsubsection{Appliance usage patterns}
\label{sec:usage_hist}

\begin{figure}[!t]
  \centering
  \includegraphics[scale=1]{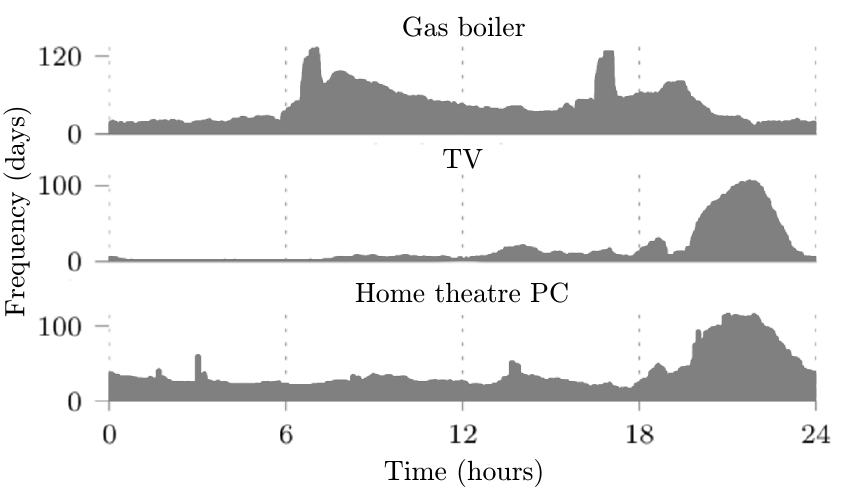}
  \caption{Daily appliance usage histograms of three appliances over 120 days from UK-DALE house 1.}
  \label{fig:daily_usage_histograms}
\end{figure} 

\noindent
Figure~\ref{fig:daily_usage_histograms} shows histograms which represent usage patterns for three appliances over
an average day, from which strong similarities between groups of appliances can be seen.  For
example, the usage patterns of the TV and Home theatre PC are very
similar because the Home theatre PC is the only video source for the TV. In contrast, the boiler has a usage pattern which occurs as a result of the household's occupancy pattern and hot water timer in mornings and evenings.

\subsubsection{Appliance correlations with weather}
\label{sec:weather_correlation}
\noindent
Previous studies have shown correlations
between temperature and heating/cooling demand in
Australia~\cite{RicharddeDear2002} and between temperature and total household demand in the USA~\cite{Kavousian2013a}.  
Such correlations could be used by a NILM system to refine its appliance usage estimates~\cite{wytock_2013}.

\begin{figure}[!t]
  \centering
  \includegraphics[scale=1]{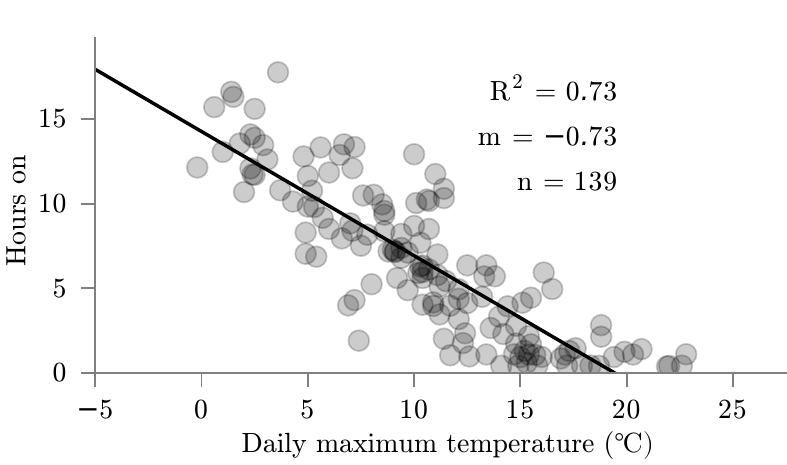} 
  \caption{Linear regression showing correlation between gas boiler
    usage and external temperature. $R^2$ denotes the coefficient of
    determination, $m$ is the gradient of the regression line and $n$
    is the number of data-points (days) used in the regression.}
  \label{fig:weather_correlations} 
\end{figure}

Figure~\ref{fig:weather_correlations} shows correlations between
boiler usage and maximum temperature (appliance data from UK-DALE house 1, temperature data from UK Met Office).  The correlation between
external maximum temperature and boiler usage is strong ($R^2=0.73$) and it is
noteworthy that the $x$-axis intercept ($\approx19\,^{\circ}\mathrm{C}$)
is approximately the set point for the boiler thermostat.


\begin{table*}
\centering
\begin{tabular}{ccccccccccc}
\hline\textbf{Data set} & \multicolumn{2}{c}{\textbf{Train time (s)}}& \multicolumn{2}{c}{\textbf{Disaggregate time (s)}} &\multicolumn{2}{c}{\textbf{NEP}}    & \multicolumn{2}{c}{\textbf{FTE}} &\multicolumn{2}{c}{\textbf{F-score}} \\ 
~ &\textbf{CO} & \textbf{FHMM} &\textbf{CO} & \textbf{FHMM} &\textbf{CO} & \textbf{FHMM} &\textbf{CO} & \textbf{FHMM}&\textbf{CO} & \textbf{FHMM} \\ \hline 
REDD &3.67 &22.81 &0.14 &1.21 &1.61 &1.35 &0.77 &0.83 &0.31 &0.31\\ 
Smart* &3.40 &46.34 &0.39 &1.85 &3.10 &2.71 &0.50 &0.66 &0.53 &0.61\\ 
Pecan Street &1.72 &2.83 &0.02 &0.12 &0.68 &0.75 &0.99 &0.87 &0.77 &0.77\\ 
AMPds &5.92 &298.49 &3.08 &22.58 &2.23 &0.96 &0.44 &0.84 &0.55 &0.71\\ 
iAWE &1.68 &8.90 &0.07 &0.38 &0.91 &0.91 &0.89 &0.89 &0.73 &0.73\\ 
UK-DALE &1.06 &11.42 &0.10 &0.52 &3.66 &3.67 &0.81 &0.80 &0.38 &0.38\\
\hline
\end{tabular}
\caption{Comparison of CO and FHMM across multiple data sets.}
\label{table:disaggregation}
\end{table*}

\subsection{Voltage Normalisation}

\noindent
Normalisation can be used to minimise the effect of voltage fluctuations in a household's aggregate power. 
Figure~\ref{fig:power_histograms} shows
histograms for both the normalised and un-normalised appliance power
consumption. Normalisation produces a noticeably tighter power
distribution for linear resistive appliances such as the toaster, although it has
little effect on constant power appliances, such as the washer dryer or LED kitchen ceiling lights. Moreover, for non-linear appliances such as the air conditioner, normalisation increases the variance in power draw. This is in conformance with work by Hart~\cite{hart_1992} which proposed a modified approach to normalisation:
\begin{equation}
\textit{Power}_{\textit{normalised}} = 
\left(\frac{\textit{Voltage}_{\textit{nominal}}}{\textit{Voltage}_{\textit{observed}}}\right)^\beta
\times \textit{Power}_{\textit{observed}}
\end{equation}
For linear appliances such as the toaster, $\beta=2$, whereas for appliances such as fridge, Hart found $\beta=0.7$. Thus, we believe the benefit of voltage normalisation is dependent on the proportion of resistive loads in a household. 

\subsection{Disaggregation Across Data Sets}

\noindent
We now compare the disaggregation results across the first house of six publicly available data sets. Again, BLUED, Tracebase and HES were not included for the same reasons as in \secref{sec:data_format}. Since all the data sets were collected over different durations, we used the first half of the samples for training and the remaining half for disaggregation across all data sets. Further, we preprocessed  the REDD, UK-DALE, Smart* and iAWE data sets to 1 minute frequency using the down-sampling filter (\secref{sec:preprocessing}) to account for different aggregate and mains data sampling frequencies and compensating for intermittent lost data packets. The small gaps in REDD, UK-DALE, SMART* and iAWE were interpolated, while the time periods where either the mains data or appliance data were missing were ignored. AMPds and the Pecan Street data did not require any preprocessing. 

Since both CO and FHMM have exponential computational complexity in the number of appliances, we model only those appliances whose total energy contribution was greater than 5\%. Across all the data sets, the appliances which contribute more than 5\% of the aggregate include HVAC appliances such as the air conditioner and electric heating, and appliances which are used throughout the day such as the fridge. We model all appliances using two states (on and off) across our analyses, although it should be noted that any number of states could be used. However, our experiments are intended to demonstrate a fair comparison of the benchmark algorithms, rather than a fully optimised version of either approach.
We compare the disaggregation performance of CO and FHMM across the following three metrics defined in \secref{sec:metrics}: (i) fraction of total energy assigned correctly (FTE), (ii) normalised error in assigned power (NEP) and (iii) F-score. These metrics were chosen because they have been used most often in prior NILM work. F-score and FTE vary between 0 and 1, while NEP can take any non-negative value. Preferable performance is indicated by a low NEP and a high FTE and F-score. The evaluation was performed on a laptop with a 2.3~GHz i7~processor and 8~GB RAM running Linux. We fixed the random seed for experiment repeatability, the details of which can be found on the project github page.


\tabref{table:disaggregation} summarises the results of the two algorithms across the six data sets. It can be observed that FHMM performance is superior to CO performance across the three metrics for REDD, Smart* and AMPds. This confirms the theoretical foundations proposed by Hart~\cite{hart_1992}; that CO is highly sensitive to small variations in the aggregate load. The FHMM approach overcomes these shortcomings by considering an associated transition probability between the different states of an appliance. However, it can be seen that CO performance is similar to FHMM performance in iAWE, Pecan Street and UK-DALE across all metrics. This is likely due to the fact that very few appliances contribute more than 5\% of the household aggregate load in the selected households in these data sets. For instance, space heating contributes very significantly (about 60\% for a single air conditioner which has a power draw of ~2.7 kW in the Pecan Street house and about 35\% across two air conditioners having a power draw of ~1.8 kW and ~1.6 kW respectively in iAWE). As a result, these appliances are easier to disaggregate by both algorithms, owing to their relatively high power demand in comparison to appliances such as electronics and lighting. In the UK-DALE house the washing machine was one of the appliances contributing more than 5\% of the household aggregate load, which brought down overall metrics across both approaches. 

Another important aspect to consider is the time required for training and disaggregation, again reported in \tabref{table:disaggregation}. These timings confirm the fact that CO is exponentially quicker than FHMM. This raises an interesting insight: in households such as the ones used from Pecan Street and iAWE in the above analysis, it may be beneficial to use CO over a FHMM owing to the reduced amount of time required for training and disaggregation, even though FHMMs are in general considered to be more powerful. It should be noted that the greater amount of time required to train and disaggregate the AMPds data is a result of the data set containing one year of data, as opposed to the Pecan Street data set which contains one week of data, as shown by \tabref{table:datasets}.

\subsection{Detailed Disaggregation Results}

\tabcolsep=0.1cm
\begin{table}[!t]
    \centering
    \begin{tabular}{ccccc}
    \hline \textbf{Appliance} & \multicolumn{2}{c}{\textbf{NEP}} & \multicolumn{2}{c}{\textbf{F-score}}\\
    ~                  & \textbf{CO}    & \textbf{FHMM}  & \textbf{CO}      & \textbf{FHMM} \\ \hline
    Air conditioner 1  & 0.3   & 0.3   & 0.9     & 0.9  \\
    Air conditioner 2  & 1.0   & 1.0   & 0.7     & 0.7  \\
    Entertainment unit & 4.2   & 4.1   & 0.3     & 0.3  \\
    Fridge             & 0.5   & 0.5   & 0.8     & 0.8  \\
    Laptop computer    & 1.7   & 1.8   & 0.3     & 0.2  \\
    Washing machine    & 130.1 & 125.1 & 0.0     & 0.0  \\
    \hline \end{tabular}
    \caption{Comparison of CO and FHMM across different appliances in iAWE data set.}
\label{table:disaggregation_iawe}
\end{table}

\begin{figure}
\includegraphics[scale=1]{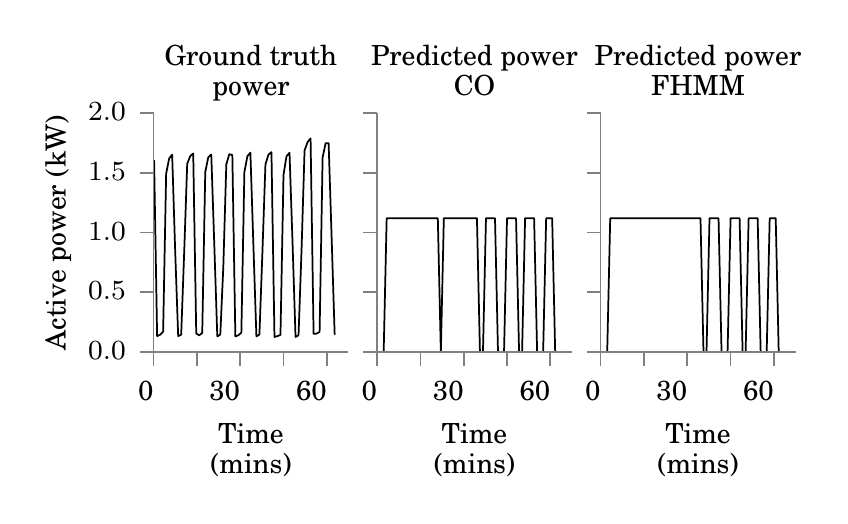} 
\caption{Predicted power (CO and FHMM) with ground truth for air conditioner 2 in the iAWE data set.}
\label{fig:ac_disaggregation} 
\end{figure}

\noindent
Having compared disaggregation results across different data sets, we now give a detailed discussion of disaggregation results across different appliances for a single house in the iAWE data set. The iAWE data set was chosen for this experiment as the authors provided metadata such as set temperature of air conditioners and other occupant patterns. \tabref{table:disaggregation_iawe} shows the disaggregation performance across the top six energy consuming appliances, in which each appliance is modelled using two states as before. It can be seen that CO and FHMM report similar performance across all appliances. We observe that the results for appliances such as the washing machine and switch mode power supply based appliances such as laptop and entertainment unit (television) are much worse when compared to HVAC loads like air conditioners across both metrics. Furthermore, prior literature shows that complex appliances such as washing machines are hard to model~\cite{barker_2013}.

We observe that the performance accuracy of air conditioner 2 is much worse than air conditioner 1. This is due to the fact that during the instrumentation, air conditioner 2 was operated at a set temperature of $26\,^{\circ}\mathrm{C}$. With an external temperature of roughly $30-35\,^{\circ}\mathrm{C}$, this air conditioner reached the set temperature quickly and turned off the compressor while still running the fan. However, air conditioner 1 was operated at $16\,^{\circ}\mathrm{C}$ and mostly had the compressor on. Thus, air conditioner 2 spent much more time in this intermediate state (compressor off, fan on) in comparison to air conditioner 1.
\figref{fig:ac_disaggregation} shows how both FHMM and CO are able to detect on and off events of air conditioner 2. Since air conditioner 2 spent a considerable amount of time in the intermediate state, the learnt two state model is less appropriate in comparison to the two state model used for air conditioner 1. This can be further seen in the figure, where we observe that both FHMM and CO learn a much lower power level of around 1.1~kW, in comparison to the rated power of around 1.6~kW. We believe that this could be corrected by learning a three state model for this air conditioner, which comes at a cost of increased training and disaggregation computational and memory requirements.

\section{Conclusions and future work}
\label{sec:conclusions}

\noindent
In this paper, we proposed NILMTK; the first open source toolkit designed to allow empirical comparisons to be made between energy disaggregation algorithms across multiple data sets. The toolkit defines a common data format, NILMTK-DF, and includes parsers from six publicly available data sets to NILMTK-DF. The toolkit further facilitates the calculation of data set statistics, diagnosing problems and mitigating them via preprocessing functions. In addition, the toolkit includes implementations of two benchmark disaggregation algorithms based on combinatorial optimisation and the factorial hidden Markov model. Finally, NILMTK includes implementations of a set of performance metrics which will enable future research to directly compare disaggregation approaches through a common set of accuracy measures. We demonstrated several analyses facilitated by NILMTK including: use of statistics functions to detect missing data, learning of appliance models from sub-metered data, comparing disaggregation algorithms across multiple data sets and breakdown algorithm performance by individual appliances.

Future work will focus upon the addition of recently proposed training and disaggregation algorithms and data sets. For instance, larger data sets such as HES could also provide additional insight into disaggregation performance.
In addition, recently proposed algorithms which do not require sub-metered power data for their unsupervised training could be compared against the current supervised algorithms.
An additional direction for future work could be the use of a semantic wiki to maintain a comprehensive, communal database of appliance metadata.
Finally, the inclusion of a household simulator (e.g.\ \cite{liang_2010}) would allow disaggregation algorithms to be evaluated in a wider variety of settings than those represented by publicly available data sets.

\section{Acknowledgements}

\noindent
Nipun Batra would like to thank TCS Research and Development for supporting him through a PhD fellowship and EMC, India for their support, and the Department of Electronic and Information Technology (Government of India) for funding the project (Grant Number DeitY/R\-\&D/ITEA/4(2)/2012). Jack Kelly thanks the EPSRC for his Doctoral Training Account and Intel for their Ph.D. Fellowship grant. Oliver Parson thanks the EPSRC for his Doctoral Prize Award. The authors thank the anonymous reviewers for their feedback, and also Denzil Correa, PhD student IIIT Delhi for his valuable comments.

\bibliographystyle{./abbrvurlmendeley.bst}

\bibliography{reference}

\clearpage
\appendix

\section{Sample code for NILMTK pipeline}
\label{app:appendix_pipeline}
\noindent
\algoref{alg:code} illustrates the NILMTK pipeline via a minimal code example.

\begin{algorithm}
\begin{verbatim}
dataset = DataSet() 

# Load the dataset
dataset.load_hdf5(DATASET_PATH)

# Load first house
building = dataset.buildings[1]

# Remove records where voltage<160
building = filter_out_implausible_values(
    building, Measurement(`voltage', `'), 160)
    
# Downsample to 1 minute
building = downsample(building, rule=`1T') 

# Choosing feature for disaggregation
DISAGG_FEATURE = Measurement(`power', `active')

# Dividing the data into train and test
train, test = train_test_split(building)

# Train on DISAGG_FEATURES using FHMM 
disaggregator = FHMM() 
disaggregator.train(train, disagg_features=
                   [DISAGG_FEATURE]) 
            
# Disaggregate
disaggregator.disaggregate(test)

# F1 score metric
f1_score = f1(disaggregator.predictions, 
            test) 
\end{verbatim}
 \caption{Example code of complete pipeline.}
 \label{alg:code}
\end{algorithm}


\section{NILMTK-DF}
\label{app:appendix_data_format}
\noindent 
We now provide the details of NILMTK-DF. \figref{fig:nilmtk-format} shows hierarchical structure used by NILMTK for modelling data sets. Each data set consists of one of more households. Each house may comprise of sensors broadly divided as utility (e.g.\ power), ambient (e.g.\ temperature) and external sensors (e.g.\ outside weather). Wherever available, we also store metadata for a household, such as area, floor, etc.
Across utilities, we focus mainly on electrical data, which is divided into mains (coming from grid), panel (circuits) and appliances. Each of these can store multiple physical measurement quantities such as active power, energy and voltage. Furthermore, we also store appliance metadata wherever available, such as rated power, details of instrumenting sensor, etc. From an implementation perspective, the lowest level in the NILMTK-DF hierarchy is stored as a Pandas DataFrame, indexed by time and with physical measured quantities such as active power as columns for each of the appliances, mains and circuits.
\begin{figure}[h]
\begin{verbatim}
dataset
|--- house_1
|   |--- ambient
|   |--- external
|   |--- metadata.json
|   |--- utility
|       |--- electricity
|       |   |--- appliances
|       |   |   |--- fridge.csv
|       |   |   |--- pump.csv
|       |   |--- circuits
|       |   |   |--- panel_1.csv
|       |   |--- mains
|       |   |   |--- mains_1.csv
|       |   |--- wiring.json
|       |--- gas
|       |--- water
|--- house_2
\end{verbatim}
\caption{NILMTK-DF format hierarchy}
\label{fig:nilmtk-format}
\end{figure}

\medskip

\section{Query Examples}
\label{app:examples}

\noindent 
We now present some of the wide range of queries supported by NILMTK.

\subsection{Across data sets}
\begin{itemize}
\item How does the daily energy consumption compare across countries?
\item How do instances of an appliance vary across countries?
\item Are there appliances which are country specific?
\end{itemize}

\subsection{Within a data set}
\begin{itemize}
\item How does the power consumption vary over seasons?
\item How does the power consumption vary between weekdays and weekend?
\item How do the power consumption of HVAC systems correlate with temperature?
\end{itemize}

\section{Functions in NILMTK}
\label{app:functions}

\noindent 
In this section we summarise the statistical (\tabref{tab:stats_functions}), diagnostic (\tabref{tab:diagnostic_functions}) and preprocessing (\tabref{tab:preprocessing_functions}) functions in NILMTK. An interested reader may refer the online documentation for updates.
\tabcolsep=0.14cm
\begin{table}

    \small
    \begin{tabular}{ll}
    \hline
    \textbf{Function} & \textbf{Definition} \\ \hline
    ON-OFF duration       & Finds the distribution of on or off\\[0.1cm]
    distribution          &durations of appliances\\[0.1cm] \hline
    Appliance usage       & Finds the temporal distribution\\[0.1cm]
    distribution          & of appliance usage\\[0.1cm] \hline
    Appliance power       & Finds the distribution of\\[0.1cm]
    distribution          & appliance power draw\\[0.1cm] \hline
    Correlation between   & Finds correlations between\\ [0.1cm]
    sensor streams        & appliances, and correlations between \\[0.1cm]
                          & appliances and other sensors\\[0.1cm] \hline  
    Find appliance        & Finds contribution of different\\[0.1cm]
    contributions         & appliances to the aggregate\\[0.1cm] \hline
    $\%$ energy sub-       & Finds the $\%$ of energy sub-metered\\[0.1cm]
    metered               &  by summing up appliance energy and \\[0.1cm]
                          &  dividing by mains energy\\[0.1cm] \hline
    $\%$ of samples      & Finds the proportion of samples\\[0.1cm]
    when energy sub-  & where the energy sub-metered is \\ [0.1cm]  
    metered greater   &above a threshold\\ 
    than threshold    &\\ \hline                                              
    \end{tabular}
    \caption{Statistical functions in NILMTK}
  \label{tab:stats_functions}
\end{table}

\tabcolsep=0.22cm
\begin{table}
    \small
    \begin{tabular}{ll}
    \hline
    \textbf{Function} & \textbf{Definition} \\ [0.1cm] \hline
    Detect gaps          & Finds gaps between readings which\\[0.1cm] 
                         & are greater than a threshold\\ [0.1cm] \hline
    Find continuous      & Finds continuous periods of data\\[0.1cm] 
    periods              & in sensor data\\[0.1cm]  \hline
    Dropout rate         & Recorded number of samples divided\\
                         & by expected number of samples\\[0.1cm] \hline
    Dropout rate         & Find the dropout rate ignoring\\[0.1cm] 
    ignoring large gaps  & large gaps\\[0.1cm]  \hline
    Uptime               & Total time for the sensor\\ [0.1cm] 
                         & which recorded readings\\ [0.1cm] \hline

    \end{tabular}
    \caption{Diagnostic functions in NILMTK}
  \label{tab:diagnostic_functions}
\end{table}

\tabcolsep=0.22cm
\begin{table}
    \small
    \begin{tabular}{ll}
    \hline
    \textbf{Function} & \textbf{Definition} \\[0.1cm] \hline
    Voltage               & Given the nominal voltage and\\[0.1cm]
    normalisation         &  observed voltage, find normalised\\[0.1cm]
                          & power draw as suggested by Hart \\[0.1cm] \hline
    Filter in top-$k$     &Filters in top-$k$ appliances by\\[0.1cm]
    appliances            &contribution to aggregate power\\[0.1cm] \hline
    Filter in appliances  &Filters in  only those appliances \\[0.1cm]
    contributing above    &which contribute more than $x$\% of\\[0.1cm]
    threshold $x$         &aggregate power\\[0.1cm] \hline
    Filter out  impl-     &Removes the readings which are\\[0.1cm]
    ausible readings      &outside of a specified range\\[0.1cm] \hline
    Filter out channels   &Removes channels which have \\[0.1cm]
    with fewer than $x$   &fewer than $x$ readings \\[0.1cm]
    readings              &\\[0.1cm] \hline
    Interpolate           &Interpolates small periods of\\[0.1cm]
                          &missing data via forward-filling\\[0.1cm] \hline
    Filter in data between& Excludes data outside the \\[0.1cm]
    start and end time    & specified start and end time\\[0.1cm] \hline
    Make common index     &Filters out times where either\\[0.1cm]
                          &mains or appliance data is absent\\[0.1cm] \hline
                                                          
    \end{tabular}
    \caption{Preprocessing functions in NILMTK}
  \label{tab:preprocessing_functions}
\end{table}

\section{Adding a new NILM algorithm}
\label{app:new_algo}
\noindent We designed NILMTK to ensure that new algorithms are easy to add. We modelled our interface on the highly successful \texttt{scikit-learn} and the R \texttt{lm} package (for Linear models). An algorithm in NILMTK needs to define the following four functions:
\begin{description}
\item[train]: Parameters of this function are the \texttt{building}, a list of \texttt{disaggregation features} (e.g.\ \texttt{[active power]} or \texttt{[active power, apparent power]}, an \texttt{aggregate stream} (e.g.\ mains) and a \texttt{sub-metered stream} (e.g.\ appliances or circuits).
The parameter style for the \texttt{train} method is similar to that of the linear regression \texttt{fit} function used in \texttt{lm}, which is as follows:

\texttt{fit <- lm(y $\sim$ x1 + x2 + x3, data = mydata)}
\item[disaggregate]: This function takes as input a \texttt{building} and disaggregates the \texttt{aggregate} feed using the appliance models learnt during training. The output is a disaggregated stream for individual appliances.
\item[import model]: This function should import a set of JSON appliance models into the NILMTK disaggregator.
\item[export model]: This function writes the learnt model to a JSON file.
\end{description}

\end{document}